\documentstyle[aps,multicol,epsf]{revtex}

\begin{document}

\title{Origin of the singular Bethe ansatz solutions for the
Heisenberg $XXZ$ spin chain}
\author{Jae Dong Noh{$^{\rm a}$}, Deok-Sun Lee{$^{\rm b}$}, and
Doochul Kim{$^{\rm b}$}}
\address{{$^{\rm a}$} Department of Physics, Inha 
University, Inchon 402-751, Korea\\
{$^{\rm b}$} School of Physics, Seoul National University, Seoul
151-742, Korea}
\maketitle

\begin{abstract}
We investigate symmetry properties of the Bethe ansatz wave functions
for the Heisenberg $XXZ$ spin chain.
The $XXZ$ Hamiltonian  commutes simultaneously with the shift operator
$T$ and the lattice inversion operator $V$ in the space of $\Omega=\pm 1$
with $\Omega$ the eigenvalue of $T$.
We show that the Bethe ansatz solutions with normalizable wave functions
cannot be the eigenstates of $T$ and $V$ with quantum number
$(\Omega,\Upsilon)=(\pm 1,\mp 1)$ where $\Upsilon$ is the eigenvalue of $V$.
Therefore the Bethe ansatz wave functions should be singular for
nondegenerate eigenstates of the Hamiltonian with quantum number
$(\Omega,\Upsilon)=(\pm 1,\mp 1)$.
It is also shown that such states exist in any nontrivial 
down-spin number sector
and that the number of them diverges exponentially with the chain length.
\end{abstract}

\begin{multicols}{2}

\section{Introduction}
The Bethe ansatz method~\cite{Bethe} has been applied to various problems
in condensed matter physics.
The Heisenberg $XXZ$ spin chain, the Hubbard model,
the Kondo model, and many other models  are solved by the Bethe
ansatz or its variants~\cite{Korepin}.
The Heisenberg $XXZ$ spin chain describes various physical systems including
interacting fermion systems.
Also, it is related to the transfer matrix 
of the six-vertex model~\cite{Baxter},
the asymmetric exclusion process,
and a Kardar-Parisi-Zhang-type growth model~\cite{Derrida}.
For the $XXZ$ Hamiltonian,
the Bethe ansatz expresses energy eigenvalues and
eigenstates as functions of quasi-particle momenta
which should satisfy a set of algebraic equations, the Bethe
ansatz equations~(BAE).
Various numerical and analytic methods have been developed 
to solve the BAE, from which
the ground state energy and low-lying excitation spectrum
are calculated~\cite{Alcaraz,Kim,BIK,BIR,WE,Noh}.
Combined with a finite-size-scaling theory, they provide useful tools to
study equilibrium and nonequilibrium critical phenomena.

When applying the Bethe ansatz method to the Heisenberg $XXZ$ spin chain,
one may doubt whether the Bethe ansatz produces the complete set of 
eigenstates.
The completeness of the Bethe ansatz is proved 
for an inhomogeneous generalization of the
$XXZ$ chain threaded by the Aharonov-Bohm flux~\cite{Langlands,Tarasov}.
It implies that the Bethe ansatz for the $XXZ$ Hamiltonian is also complete 
in a general sense. One may obtain the eigenstate of the $XXZ$ Hamiltonian
from that of the generalized Hamiltonian by taking a limit where
the inhomogeneity and the flux vanish.

On the other hand, it has also been investigated for a long time whether
the Bethe ansatz produces all eigenstates of the $XXZ$ Hamiltonian without
the help of the limiting procedure.
There have been the arguments for the completeness 
assuming a string conjecture.
For the $XXX$ chain, which  has isotropic spin-spin interaction,
Takahashi~\cite{Takahashi} showed that the number of the states
constructed from the Bethe ansatz using the string conjecture
is equal to the total number of the spin states.
The string conjecture was also adopted to prove the completeness of the
Bethe ansatz for a generalized $XXZ$ chain~\cite{Kirillov} and
the Hubbard model~\cite{Essler91}.
Later, the string conjecture was shown to be invalid by
Essler {\em et al.}~\cite{Essler}, who analyzed the algebraic
Bethe ansatz equations in the two down-spin sector.
They found that some solutions expected from the string conjecture
are missing and that there exist solutions violating the string
conjecture.
Nevertheless, they concluded that the Bethe ansatz is complete
at least in the two down-spin sector because the numbers of the former
and the latter are the same.

However one of the solutions found in Ref.~\cite{Essler} is {\em
singular}; the corresponding Bethe ansatz wave function is not well-defined.
In fact the state $|\Psi\rangle \ = \sum_l (-1)^l |l,l+1\rangle$, which
is an eigenstate of the $XXZ$ Hamiltonian, is not produced from the {\em
regular}~(non-singular) solutions of the Bethe ansatz~(see below for the
definition of the state $|l,l+1\rangle$).
Quasi-particle momenta corresponding to the state are
divergent~\cite{Bethe,Siddharthan,Wal} and the Bethe ansatz wave function is
ill-defined.
An eigenstate that is not produced from the regular solutions of the
Bethe ansatz will be referred to as a singular state.
Siddharthan~\cite{Siddharthan} reported that
there also exist the singular states in the three down-spin sector.
It was claimed that a symmetry property might be
important since the singular states have a definite symmetry property.

The singular state does not imply the incompleteness of the Bethe ansatz for
the $XXZ$ Hamiltonian. 
One may obtain the wave functions for the singular states of the
$XXZ$ Hamiltonian from those of the generalized
Hamiltonian by taking the limit where the inhomogeneity and the flux
vanish~(see, for example, Ref.~\cite{Siddharthan}).
However it should be examined whether the singular states
are relevant to the involved physical quantities of the Heisenberg
$XXZ$ spin chain since, 
if so, one must first consider the generalized model and take the suitable
limit~\cite{Langlands,Tarasov,Siddharthan}.
Therefore, it is helpful to answer the  question why there appear
the singular states and 
how many there are in general down-spin number sectors. 
In this paper, we investigate the symmetry properties of the $XXZ$
Hamiltonian and the Bethe ansatz wave functions
to show that the regular solution of the Bethe ansatz
with well-defined wave functions does not contain a certain class
of eigenstates. Those states are the singular states.
We also estimate the number of the singular states approximately
at each sector with fixed number of down spins and find that
the number of them diverges exponentially with the chain length.

\section{Bethe states and symmetry operations}
The Heisenberg $XXZ$ spin chain of length $N$ has the Hamiltonian
\begin{equation}
H = - \frac{1}{2}\sum_{l=1}^N \left\{ \sigma_l^+ \sigma_{l+1}^-
+  \sigma_l^- \sigma_{l+1}^+ + \frac{\Delta}{2}
( \sigma_l^z \sigma_{l+1}^z - 1 ) \right\} \ ,
\end{equation}
where $\sigma_l^{x,y,z}$ are the
Pauli matrices at site $l$,
$\sigma_l^{\pm} = (\sigma_l^x \pm i \sigma_l^y)/2$. The periodic
boundary condition is imposed, $\sigma_{N+1}^{x,y,z} = \sigma_1^{x,y,z}$.
We assume $N$ is even unless stated otherwise.
It reduces to the ferromagnetic (antiferromagnetic)
$XXX$ chain in the case of $\Delta = 1$ ($\Delta = -1$).
Since the Hamiltonian commutes with the magnetization
$M=\sum_l \sigma_l^z$, one can work in the $Q$ down-spin sector 
in the diagonal basis of $\sigma^z$.
A state in the $Q$ down-spin sector is spanned by $_NC_Q$ state vectors
$|x_1,x_2,\ldots,x_Q\rangle$ with $1\leq x_i\leq N$~($x_i<x_j$
for $i<j$) denoting the position of the $i$th down spin.

The eigenstate of $H$ can be written as
$|\Psi\rangle = \sum_{x_1<\cdots<x_Q} \psi(x_1,\ldots,x_Q)
|x_1,\ldots,x_Q\rangle$.
According to the Bethe ansatz, the wave function is given in terms of
quasi-particle momenta $k_m~(m=1,\ldots,Q)$ by
\begin{equation}\label{psi}
\psi(x_1,\ldots,x_Q) = \sum_{P} A(P) \exp\left( i \sum_{m=1}^Q k_{P_m} x_m
\right) \ ,
\end{equation}
where the sum is over all permutations
$P$ of integers $\{1,2,\ldots,Q\}$. The quasi-particle momenta are complex
numbers with $-\pi< {\rm Re }~{k}\leq \pi$. The amplitudes $A$'s
for different permutations $P$ and $P'$, which are identical except for
two neighboring integers such that
$P_j=P'_{j+1}=m, P_{j+1}=P'_j=l$, and $P_i=P'_i$
for $i\neq j,j+1$, are related as
\begin{equation}\label{perm}
\frac{A(P)}{A(P')} =
-\frac{e^{i(k_m+k_l)} - 2 \Delta e^{i k_m} + 1}{e^{i(k_m+k_l)} -
2 \Delta e^{i k_l} + 1} \ .
\label{amplitude}
\end{equation}
The periodic boundary condition implies the Bethe ansatz
equations~(BAE) for the quasi-particle momenta:
\begin{equation}\label{BAE}
e^{iNk_m} = (-1)^{Q-1} \prod_{l\neq m} \frac{e^{i(k_m+k_l)}-2\Delta
e^{ik_m}+1}{ e^{i(k_m+k_l)}-2\Delta e^{ik_l} +1} \quad .
\end{equation}
The state constructed from the Bethe ansatz is called the Bethe state.
It has the energy $E = \sum_{m=1}^Q (\Delta-\cos k_m)$.

In the special case of $\Delta=0$, where the BAE becomes
$e^{ik_mN}=(-1)^{Q-1}$, the Bethe ansatz produces all eigenstates of $H$.
A solution is obtained by selecting $Q$ different values among the $N$
values of $(\frac{2\pi}{N})\times$ integers~(half-integers)
for odd~(even) $Q$. There are ${}_NC_Q$ different
ways for each $Q$, which is equal to the possible spin states.
Naively one may think that the solutions at $\Delta=0$ evolve
continuously as $\Delta$ is turned on.
However it will turn out to be false.

The $XXZ$ Hamiltonian under the periodic boundary condition
commutes with the shift operator $T$,
which shifts the position of each down spin to the left by one unit.
By construction, the Bethe state is
the eigenstate of $T$ with the eigenvalue $\Omega=[\prod_{m=1}^Q e^{ik_m}]$.
Since $T^N$ is the identity operator, $\Omega^N$ should be unity.
The Hamiltonian also commutes with the lattice inversion operator $V$
defined by $V |\{x_i\}\rangle = |\{y_i\}\rangle$ with $y_i = N-x_{Q-i+1}+1$.
Since $V^2$ is the identity operator, the eigenvalue of $V$,
which will be denoted by $\Upsilon$, takes the value of $\pm 1$.
The inversion operator does not commute with the shift operator in
general. But it is easy to show that they commute with each other
in the subspace of $\Omega=\pm 1$; $[T,V] | \Psi \rangle = 0 $ if
$ T | \Psi \rangle = \pm | \Psi \rangle $.
Then the eigenstate of $H$ can be made as the simultaneous eigenstate of $T$
and $V$ in the subspace.

The Bethe state with $\Omega=\pm 1$ is not the eigenstate of $V$
necessarily. If a Bethe state $|\Psi\rangle$ with $\Omega=\pm 1$
with well-defined wave function is a simultaneous eigenstate of
$V$~($V|\Psi\rangle = \Upsilon | \Phi\rangle$),
then one can show that $\Upsilon$ should be equal to $\Omega$ :
\begin{equation}\label{lemma}
\Upsilon = \Omega \ .
\end{equation}
It is proved as follows.
Suppose that $|\Psi\rangle$ is a Bethe state in the $Q$  down-spin sector
with the quasi-particle momenta $\{k_1,k_2,\ldots,k_Q\}$ satisfying
Eq.~(\ref{BAE}). The wave function is given as in Eq.~(\ref{psi}).
When one applies the lattice inversion operator $V$ to the state
$|\Psi \rangle$, the wave function of the state $V|\Psi\rangle = \sum
\widetilde\psi(x_1, \ldots,x_Q) | x_1,\ldots,x_Q\rangle$ is given by
\begin{eqnarray}
\widetilde{\psi}(x_1,\ldots,x_Q)
&=& \psi(N-x_Q +1,\cdots,N-x_1+1) \nonumber \\
&=& \Omega \sum_P \widetilde{A}(P)
\exp \left[ i \sum_m (-k_{P_m}) x_m \right]  \ ,
\label{Vwf}
\end{eqnarray}
where
$\widetilde{A}(P)=A(P \bar{P})$ with a parity permutation
$\bar{P}$ which maps $\{1,2,\cdots,Q\}$ to $\{Q,\cdots,2,1\}$,
and we used that $[\prod_{m=1}^Q e^{i k_m}] = \Omega$ and $\Omega^N=1$.
One can easily verify that $\{-k_m\}$ satisfies the BAE in
Eq.~(\ref{BAE}) and that the amplitude $\widetilde{A}(P)$ satisfies
Eq.~(\ref{perm}) with $\{k_m\}$ replaced by $\{-k_m\}$. This shows
that $V|\Psi\rangle$ is a Bethe state with the quasi-particle momenta
$\{-k_1,-k_2,\ldots,-k_Q\}$.
Therefore a Bethe state could be an eigenstate of $V$ only when
the two sets $\{k_1,k_2,\ldots,k_Q\}$ and $\{-k_1,-k_2,\ldots,-k_Q\}$
are identical.

We rewrite the wave function in Eq.~(\ref{Vwf}),
introducing the permutation $\widetilde{P}$ defined by the relation
$-k_l = k_{\widetilde{P}_l}$, as
\begin{eqnarray}
\tilde{\psi}(\{x_m\})
&=&\Omega\sum_P \widetilde{A}(P)
\exp\left[i\sum_m k_{\widetilde{P}P_m}x_m\right] \nonumber \\
&=&\Omega\sum_P A(\tilde{P}P\bar{P})\exp\left[i\sum_m k_{P_m}x_m\right] \ .
\end{eqnarray}
Comparing it with Eq.~(\ref{psi}), one can find $\Upsilon$ from the ratio of
$A(P)$ and $\Omega A(\tilde{P}P\bar{P})$ for a certain permutation $P$,
e.g., the identity permutation $I$. 
If the amplitude $A$ is nonzero and finite, $\Upsilon$ is given by
\begin{equation}
\Upsilon = \Omega A(\tilde{P}\bar{P}) / A(I).
\end{equation}

Note that the set $\{k_m\}$  is equal to $\{-k_m\}$.
On one hand, unless both $0$ and $\pi$ are present in the set $\{k_m\}$, 
one can rearrange the momenta in such a way that the permutation 
$\widetilde{P}$ and the parity permutation $\bar{P}$ are the same.
Then one has $A(\tilde{P}\bar{P}) = A(\bar{P}^2) = A(I)$ and hence
$\Upsilon = \Omega$. On the other hand, if both $0$ and $\pi$ are present 
in $\{k_m\}$,
$Q$ should be even and one can rearrange the momenta in the following way:
$$
\{k_1,\ldots,k_{Q/2-1},k_{Q/2},k_{Q/2+1},
-k_{Q/2-1},\ldots,-k_1\} \ ,
$$
with $k_{Q/2}=0$ and $k_{Q/2+1}=\pi$.
With this arrangement, the permutation $\widetilde{P}\bar{P}$ maps
$Q/2$ to $Q/2+1$, $Q/2+1$ to $Q/2$, and $j\neq Q/2,Q/2+1$ to $j$.
Then, using Eq.~(\ref{perm}), the ratio between $A(I)$ and
$A(\tilde{P}\bar{P})$ is given by
\begin{equation}
\frac{A(I)}{A(\tilde{P}\bar{P})}=
-\frac{e^{ik_{Q/2+1}+ik_{Q/2}}-2\Delta e^{ik_{Q/2}}+1}
{e^{ik_{Q/2+1}+ik_{Q/2}}-2\Delta e^{ik_{Q/2+1}}+1} = 1 \ .
\end{equation}
Therefore we find that $V|\Psi\rangle=\Omega |\Psi\rangle$ if
$|\Psi\rangle$ is an eigenstate of $V$.
This proves Eq.~(\ref{lemma}).

\section{Perturbative construction of singular states}
Equation~(\ref{lemma}) of previous section shows that the Bethe state with 
well-defined wave function cannot be the simultaneous
eigenstate of $T$ and $V$ with
$(\Omega,\Upsilon)=(\pm 1,\mp 1)$.
But there exist many eigenstates of $H$ with such symmetry.
Suppose an eigenstate of $H$ possesses the quantum numbers
$(\Omega,\Upsilon)=(1,-1)$. The state may be degenerate~(having the same
energy) with another state which has $\Upsilon=+1$ in the same $\Omega$
sector.
In the degenerate case, the Bethe state may be a  linear superposition of the
degenerate pair, which does not lead to the contradiction to
Eq.~(\ref{lemma}).
On the other hand, if the eigenstate does not have the degenerate partner
with $\Upsilon=+1$ in the same $\Omega$ sector,
such a state cannot be a Bethe state, hence is absent in the regular 
Bethe ansatz solutions. In the remaining part, we show that there exist 
nondegenerate states 
with $(\Omega,\Upsilon)=(\pm 1,\mp 1)$ for nonzero $\Delta$ in the general 
$Q$ down-spin sector. 

It is convenient to work with the second quantized form of the
Hamiltonian.
Under the Jordan-Wigner and Fourier transformations, defined as
$\sigma_l^+ = a_l \exp[i\pi\sum_{j=1}^{l-1} a_j^\dagger a_j]$
and $a_l = \sum_p a_p e^{ipl}/\sqrt{N}$, respectively,
the $XXZ$ Hamiltonian in the $Q$ down-spin sector reads as 
$H= H_0 + H_1$ where $H_0 = -\sum_p (\cos p-\Delta) \ n_p $ and
$$
H_1 = -\frac{\Delta}{N} \sum_{p_1+p_2=p_3+p_4} e^{-i(p_1-p_4)}
a_{p_1}^\dagger a_{p_2}^\dagger a_{p_3} a_{p_4}.
$$
Here $a_p$ and $a_p^\dagger$ are anticommuting fermion operators with the
bare momentum $-\pi<p\leq \pi$ and $n_p=a_p^\dagger a_p$ is the number
operator with $\sum_p n_p = Q$.
The momentum $p$ takes the real value of $({2\pi}/{N}) \times$
integer~(half-integer) for odd~(even) $Q$. 

\begin{figure}
\centerline{\epsfxsize=8.7cm \epsfbox{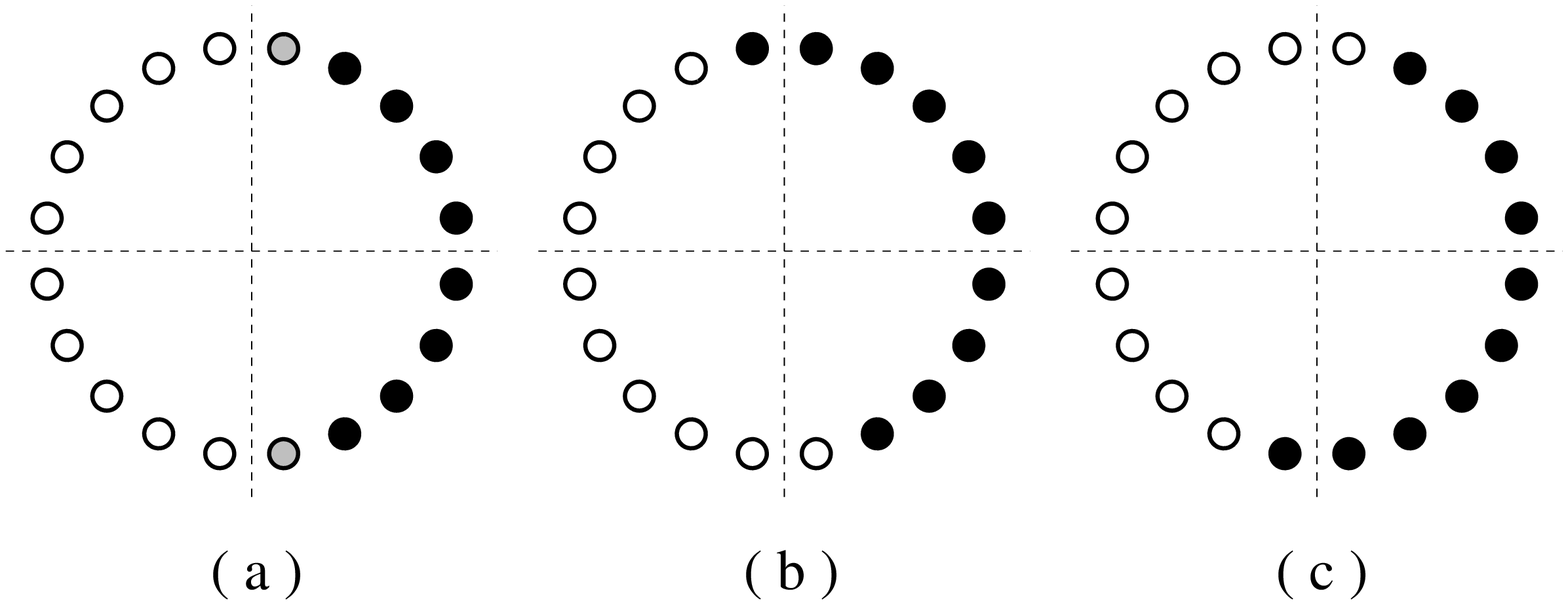}}
\caption{An illustration of sets $S_{Q-2}$ and $R$, and a state in
Eq.~(\ref{SQ2}) with $N=20$ and $Q=10$.
The bare momenta $p$ are denoted by equi-spaced circles in a complex
$z\equiv e^{ip}$ plane.  The horizontal~(vertical) dashed lines stand for
the real~(imaginary) $z$ axis.
(a) The elements of the set $S_{Q-2}~(R)$ are denoted by black~(gray)
circles. In (b) and (c) black circles denote the occupied momenta in
Eq.~(\ref{SQ2}).}
\label{circle}
\end{figure}

Consider a symmetric set $S_{Q-2}$ which consists of $(Q-2)$ bare momenta
satisfying $\sum_{p\in S_{Q-2}}p = 0$~(mod $2\pi$).
A set $S$ is symmetric if it contains $-p$ for all $p\in S$.
For a given $S_{Q-2}$, $R$ is defined as the set of all momenta in the interval
$-{\pi}/{2}<p<{\pi}/{2}$ except
$p'$ or $\pi-p'$~(mod $2\pi$) for $p'\in S_{Q-2}$.
Note that $R$ is also a symmetric set.
Then, at $\Delta=0$, the following states
\begin{equation}\label{SQ2}
|p_\alpha;S_{Q-2}\rangle \equiv a^\dagger_{p_\alpha}
a^\dagger_{\pi-p_\alpha} \prod_{p\in
S_{Q-2}} a^\dagger_p |0\rangle
\end{equation}
with $p_\alpha \in R$ are degenerate eigenstates of $H$
in the $Q$ down-spin sector having $\Omega=-1$.
Here $|0\rangle$ denotes the vacuum, i.e., $a_p |0\rangle = 0$ for all $p$.
They are not the eigenstates of $V$. One can easily show that
\begin{equation}\label{VpS}
V | p_\alpha;S_{Q-2}\rangle = - \Omega | -p_\alpha; S_{Q-2}\rangle \ .
\end{equation}
Figure~\ref{circle} illustrates an example of the set $S_{Q-2}$ and
corresponding $R$ with the states in Eq.~(\ref{SQ2}) for $N=20$ and $Q=10$.

As $\Delta$ is turned on, $H_1$ generates the overlap between the degenerate
states, which may cause the degeneracy to split.
To see the degeneracy splitting, we treat $H_1$ as a perturbation
and apply the degenerate perturbation theory.
In this scheme, the leading order correction to the energy is
given by the eigenvalue of the perturbation matrix
$(H_1)_{p_\alpha,p_\beta} \equiv \langle{p_\alpha;S_{Q-2}}| H_1 |
{p_\beta;S_{Q-2}}\rangle$ with $p_\alpha, p_\beta\in R$.
The eigenstates are given by the eigenvectors of the perturbation matrix. 
A straightforward calculation shows that
the matrix element is given by
$$ (H_1)_{p_\alpha,p_\beta} =
-\frac{4 \Delta}{N} (\cos{p_\alpha}) (\cos{p_\beta}) \ ,$$
where we neglect a constant diagonal element.
It has the nondegenerate eigenvector
\begin{equation}
|\Psi;S_{Q-2}\rangle = \sum_{p_\alpha\in R}
\cos{p_\alpha} |p_\alpha;S_{Q-2}\rangle \ ,
\end{equation}
with the eigenvalue $-(4\Delta/N)\sum_{p_\alpha \in R} \cos^2 p_\alpha$.
The other states remain to be degenerate with their common eigenvalue $0$.
If one applies $V$ to the state $|\Psi;S_{Q-2}\rangle$ and
uses Eq.~(\ref{VpS}),
one can see that the nondegenerate state is the eigenstate
of $V$ with $\Upsilon=+1$ while $\Omega=-1$.
Therefore, it cannot be produced from the regular solution of the Bethe
ansatz by Eq.~(\ref{lemma}).

The remaining degenerate states may split further in the higher order
perturbations.
So the number of the nondegenerate states with $(\Omega,\Upsilon)=(-1,+1)$
is equal to or greater than the number of different realizations of 
the set $S_{Q-2}$.
For even $Q$, $S_{Q-2}$ consists of  $(Q-2)/2$ values of $p$
selected randomly in the interval $0<p<\pi$ and their negative partners.
For odd $Q$, $S_{Q-2}$ consists of $p=0$, $(Q-3)/2$ $p$'s
randomly-selected in the interval $0<p<\pi$, and their negative partners.
So, there are at least $_{N/2}C_{(Q-2)/2}$~($_{N/2}C_{(Q-3)/2}$)
states which have $(\Omega,\Upsilon)=(-1,+1)$ and
they are those states which cannot be produced
from the Bethe ansatz in the even~(odd)
$Q$ down-spin sector. If we sum over all $Q$, the number of such states
is of the order of $2^{N/2}$.

In the two down-spin sector, the perturbation calculation becomes exact since
the states in Eq.~(\ref{SQ2}) with the empty set $S_{Q-2}$ span
the whole two down-spin sector with $\Omega=-1$. So the state
$$ \sum_{-\frac{\pi}{2} < p < \frac{\pi}{2}} (\cos{p})\ a_p^\dagger a_{\pi-p}^\dagger |0\rangle $$
is the eigenstate of $H$ at all $\Delta$, which cannot be produced from the
Bethe ansatz. If one performs the inverse Jordan-Wigner and Fourier
transformations, the state can be written as $\sum_{l} (-1)^l |l,l+1\rangle$.
This is the example of the singular states mentioned in~\cite{Siddharthan}.

Consider also the states
$$
a^\dagger_{p_\alpha}a ^\dagger_{p_\beta} a^\dagger_{\pi-p_\alpha}
a^\dagger_{\pi-p_\beta} \prod_{p\in S_{Q-4}} a^\dagger_p |0\rangle
$$
with $p_\alpha,p_\beta\in R$, constructed from a symmetric set
$S_{Q-4}$ consisting of $(Q-4)$ momenta satisfying
$\sum_{p\in S_{Q-4}}p=0$~(mod $2\pi$) and the set $R$ of all momenta in the
interval $-\frac{\pi}{2}<p<\frac{\pi}{2}$ except
$p'$ or $\pi-p'$~(mod $2\pi$) for $p' \in S_{Q-4}$.
They are degenerate eigenstates of $H$ at $\Delta=0$ with $\Omega=+1$.
The same perturbation analysis will show that the degeneracy splits for
finite $\Delta$ and there appear nondegenerate states
with $\Upsilon = -1$.
Those states cannot be produced by the Bethe ansatz from
Eq.~(\ref{lemma}) either.
There are ${}_{N/2}C_{(Q-4)/2}$~(${}_{N/2}C_{(Q-5)/2}$)
different ways in choosing the set $S_{Q-4}$ for even~(odd) $Q$.
For each set, there exists at least one state with $(\Omega,\Upsilon)=(+1,-1)$.
So the total number of such states is again of order of $2^{N/2}$.

The same analysis can be applied to the states from the symmetric sets
$S_{Q-6}, S_{Q-8},\ldots$, with $\sum_{p\in S}p=0$~(mod $2\pi$) and
symmetric sets $S_{Q-1}, S_{Q-3},\ldots$ containing $p=\pi$ and
satisfying $\sum_{p\in S}=\pi$~(mod $2\pi$). For each case, there are at
least a number of order of $2^{N/2}$ singular states.

\section{Numerical Results}
The state counting discussed in the previous section is
not exact. States from different $S$ sets may be degenerate at
$\Delta=0$. In that case perturbation calculations should be carried out
in an enlarged space and it is not guaranteed that each $S$ set generates
at least one singular state. And the singular states with
$(\Omega,\Upsilon)=(\pm 1,\mp 1)$ may become degenerate accidentally with a
state with $(\Omega,\Upsilon)=(\pm 1,\pm 1)$ for a particular value of
$\Delta$. Then the Bethe ansatz may produce the linear superpositions of
them. We believe that those effects do not affect the leading order of
magnitude of the total number of the singular states.

\begin{figure}
\centerline{\epsfxsize=8.7cm \epsfbox{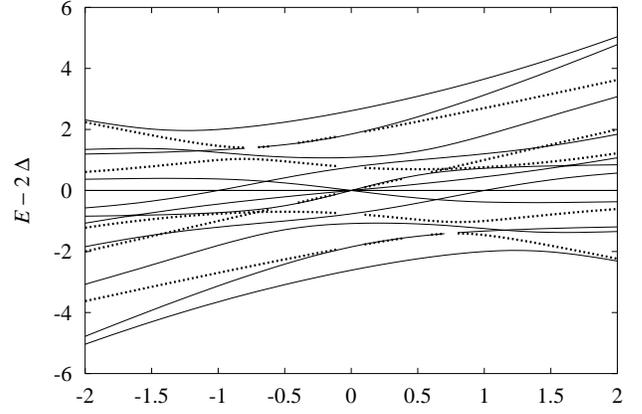}}
\caption{Energy vs. $\Delta$ with $\Omega=\pm 1$ of the $XXZ$ Hamiltonian in the
four down-spin sector with the chain length $N=8$.
The dotted lines represent the singular
states, nondegenerate eigenstates with quantum number 
$(\Omega,\Upsilon)=(\pm 1,\mp 1)$.  }\label{level}
\end{figure}

We checked this by diagonalizing the $XXZ$ Hamiltonian exactly with
$N=6,8,\ldots,16$. The symmetry property of each
eigenstate is checked and the nondegenerate states with
$(\Omega,\Upsilon)=(+1,-1)$ and ($-1,+1$) are identified as the singular
states. Figure~\ref{level} shows the energy level, as a function of $\Delta$,
of each state with $\Omega=\pm 1$ for
$N=8$ in the four down-spin sector. The singular states are denoted by
dotted lines. Note that each singular state evolves from the degenerate states
at $\Delta =0$ as the degeneracy is lifted for nonzero $\Delta$ .
At $\Delta=\pm \frac{1}{2}$, many singular states become degenerate with
non-singular states and
they are not counted as the singular states.
In Table 1, we show the total number of the singular states at $\Delta=0.2,
0.5$, and $1.0$. Due to the accidental degeneracy the number varies with
$\Delta$. It is confirmed that the number increases approximately as $2^{N/2}$
at the three values of $\Delta$.

\section{Discussion and Summary}
The singular behaviors of some Bethe ansatz solutions have been noticed
for the Heisenberg $XXZ$ model~(see, e.g., \cite{Bethe,Alcaraz,Siddharthan}).
So the generalizations to the inhomogeneous $XXZ$ model with Aharonov-Bohm
flux were considered and their Bethe ansatz solutions were shown to be
complete~\cite{Langlands,Tarasov}.
In those cases the Hamiltonian does not possess the lattice inversion
symmetry and the present analysis is not applicable. 

The singular states include low-lying states in the half-filling case,
$Q=N/2$. Figures 1 (b) and (c) illustrate degenerate first
excited states at $\Delta=0$ in the $\Omega=+1$ sector.
They are not the eigenstates of the
inversion operator $V$. For nonzero $\Delta$ the degeneracy is removed,
and the states evolve into the eigenstates of $V$ with $\Upsilon=\pm 1$.
Therefore the Bethe ansatz fails to produce one of them with
$\Upsilon=-1$.
Those states are important in identifying the operator content of
the $XXZ$ model in the context of
the conformal field theory~(see \cite{Alcaraz} and references therein).
They have the scaling dimension $x_{n,m}$ with $n=0$ and $m=\pm 1$.
To recover the whole operator content from the Bethe ansatz, one should
introduce the Aharonov-Bohm flux and take the limit
where the flux vanishes as done in~\cite{Alcaraz}.

So far we have assumed $N$ is even.
For odd $N$, one cannot find a pair of bare momenta $p$ and $p'$
satisfying $p+p'=\pi$ and the previous analysis cannot be applied.
So the above arguments are not valid for odd $N$, though
Eq.~(\ref{lemma}) holds good.
We have performed the same numerical analysis for odd $N\leq 15$ and
found that the nondegenerate states with $(\Omega,\Upsilon)=(\pm 1,\mp 1)$
do not exist.

In summary, we have shown that the symmetry properties of the Hamiltonian 
and the Bethe wave functions prevent the regular solutions of the Bethe ansatz 
from containing some eigenstates. 
For the fixed number of down spins, the $XXZ$ Hamiltonian
can be simultaneously diagonalized with
the shift operator $T$ and the lattice inversion operator $V$ in
the subspace where the eigenvalue of $T$, $\Omega$, is $\pm1$,
and each simultaneous eigenstate has one of the following four
pairs of eigenvalues,
$(\Omega,\Upsilon) = (1,1),(-1,-1),(1,-1),(-1,1)$.
But there is no  regular Bethe ansatz wave function with 
$(\Omega,\Upsilon) = (1,-1),(-1,1)$, which leads to the appearance of
the singular Bethe states.
We estimated the number of the singular states as of
$O(2^{N/2})$ for even $N$.
The number of the singular states is exponentially large
though the fraction to the total number of the states, $2^N$, is
vanishingly small.
Moreover, the singular states include low-lying states in the
half-filling case where the number of down spins is $N/2$.

\acknowledgments
This work was supported in part by the BK21 Project 
of Ministry of Education, Korea.

\begin{table}
\caption{
The number of the nondegenerate states
with  $(\Omega,\Upsilon)=(\pm 1,\mp 1).$}
\begin{tabular}{cccc} 
$N$ & $\Delta=0.2$ & $\Delta = 0.5$ & $\Delta=1.0$  \\ \hline 
8   & 12  & 7   &12  \\
10  & 25  & 15  & 20  \\
12  & 64  & 29  & 59  \\
14  & 152 & 61  & 152 \\
16  & 339 & 131 & 319  
\end{tabular}
\end{table}

\end{multicols}
\end{document}